\documentclass[12pt]{article}
\usepackage[pdftex]{graphicx}
\usepackage{amssymb,amsfonts,amsmath}
\usepackage[font=small, format=hang]{caption}
\usepackage{hhline}
\usepackage{lineno}
\usepackage{fancyhdr}
\usepackage{calc}
\usepackage{url}
\usepackage{amsthm}
\begin{document}

\begin{titlepage}

{\sc \uppercase{\textbf{Spatial effects on species persistence and implications for biodiversity}}}

\vspace{3truecm}
\sc {Enrico Bertuzzo$^1$, Samir Suweis$^1$,  Lorenzo Mari$^1$, Amos Maritan$^2$, Ignacio Rodriguez-Iturbe$^3$ and Andrea Rinaldo$^{1,4}$}

\vspace{3truecm}
\centerline{\hbox to 5.5truecm{\hrulefill}}
\medskip
%
{\sc$^1=$\'Ecole Polytechnique F\'ed\'erale Lausanne, 1015 Lausanne,Switzerland}\\
{\sc$^2=$Department of Physics, University of Padova, 35131 Padova, Italy}\\
{\sc$^3=$Department of Civil
and Environmental Engineering, Princeton University, Princeton, NJ
08544, USA}\\
{\sc$^4=$Dipartimento IMAGE, Universit\'a di
Padova, 35131 Padova, Italy.}\\

%
%
%

\title{Spatial effects on species persistence and implications for biodiversity}
\author{Enrico Bertuzzo\affil{1}{Laboratory of Ecohydrology,
\'Ecole Polytechnique F\'ed\'erale Lausanne, 1015 Lausanne,
Switzerland}, Samir Suweis\affil{1}{}, Lorenzo Mari\affil{1}{}, Amos Maritan\affil{2}{Department of Physics, University of Padova, 35131 Padova, Italy}, Ignacio Rodriguez-Iturbe\affil{3}{Department of Civil
and Environmental Engineering, Princeton University, Princeton, NJ
08544, USA}, \and Andrea
Rinaldo\thanks{To whom correspondence should be addressed. E-mail:
andrea.rinaldo@epfl.ch}\affil{1}{}\affil{4}{Dipartimento IMAGE, Universit\'a di
Padova, 35131 Padova, Italy.}}


\end{titlepage}

\clearpage

\begin{abstract}
Natural ecosystems are characterized by striking diversity of form
and functions and yet exhibit deep symmetries emerging across
scales of space, time and organizational complexity. Species-area
relationships and species-abundance distributions are examples of
emerging patterns irrespective of the details of the underlying
ecosystem functions. Here we present empirical and theoretical
evidence for a new macroecological pattern related to the
distributions of local species persistence times, defined as the timespans
between local colonizations and extinctions in a given geographic
region. Empirical distributions pertaining to two different taxa,
breeding birds and herbaceous plants, analyzed in a new framework
that accounts for the finiteness of the observational period,
exhibit power-law scaling limited by a cut-off determined by the
rate of emergence of new species. In spite of the differences
between taxa and spatial scales of analysis, the scaling
exponents are statistically indistinguishable from each other and significantly
different from those predicted by existing models. We
theoretically investigate how the scaling features depend on the
structure of the spatial interaction network and show that the
empirical scaling exponents are reproduced once a two-dimensional
isotropic texture is used, regardless of the details of the ecological
interactions. The framework
developed here also allows to link the cut-off timescale with the
spatial scale of analysis, and the persistence-time distribution to
the species-area relationship. We conclude that the inherent
coherence obtained between spatial and temporal macroecological
patterns points at a seemingly general feature of the dynamical
evolution of ecosystems.
 \end{abstract}

\emph{keywords}: \;\, persistence times,macroecology,spatial ecology,biogeography
\clearpage
\newpage

\section{Introduction}
\begin{figure}
  \includegraphics[width=29pc]{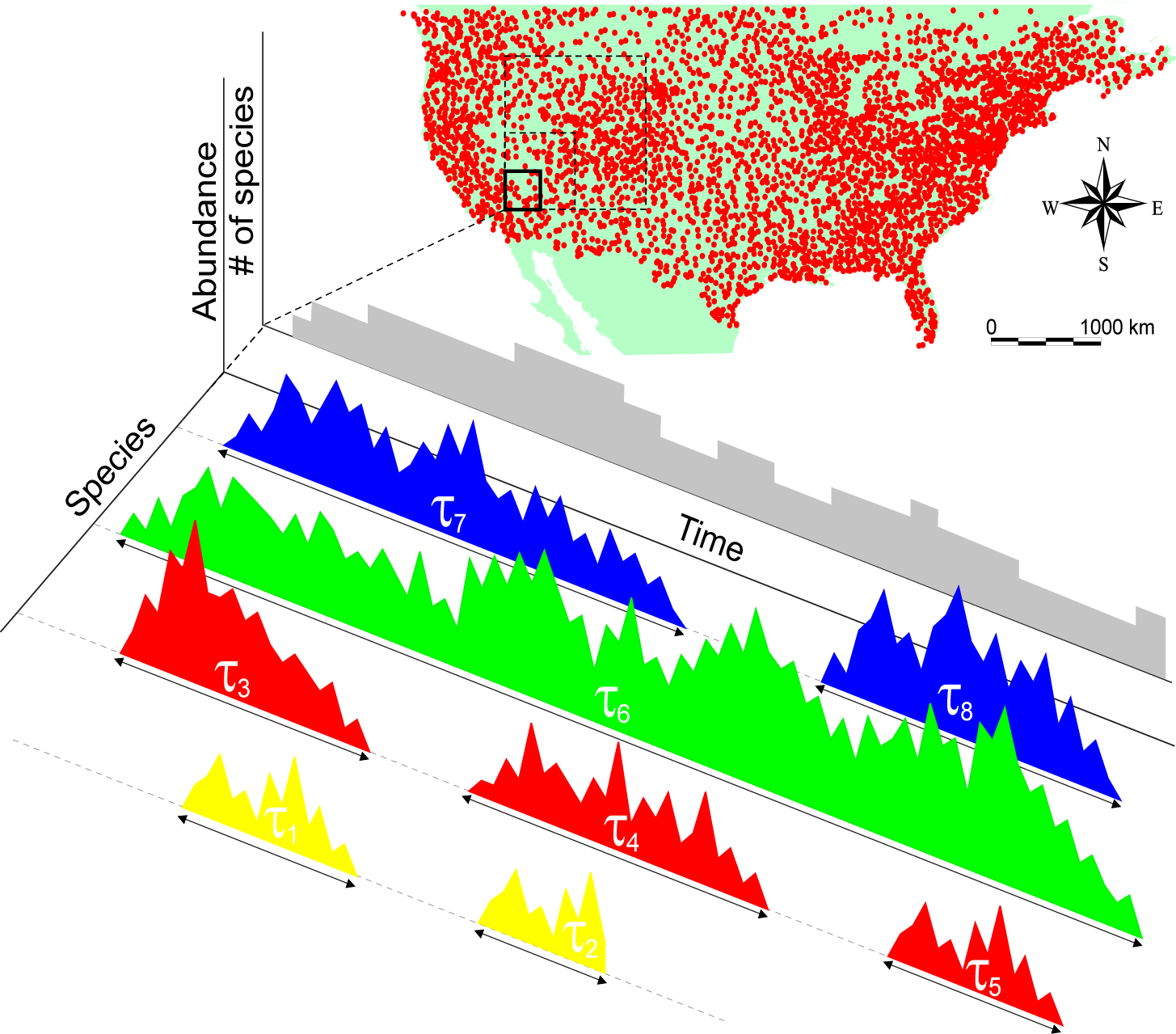}\\
\caption{}\label{fig1}
\end{figure}

Understanding local extinction processes has gained urgency as the
number of threatened species increases throughout the world
because of factors like habitat destruction or climate change~\cite{diamond1989,brown1995,Thomas2004,svenning2008,May2010},
but a synthesis of theory and empirical evidence accounting for
the relevant ecological dynamics is lacking. In this context, we address here the
study of persistence times of trophically equivalent co-occurring species
in relation to the spatial scale of observation. The persistence time
$\tau$ of a species within a geographic region is defined as
the time incurred between its emergence and its local extinction
(see~\cite{keitt1998,pigolotti2005} and Fig.~\ref{fig1}). At a
local scale, persistence times are largely controlled by ecological
processes operating at short timescales (e.g. population dynamics, dispersal,
immigration, contraction/expansion of species geographic ranges)
as local extinctions are dynamically balanced by colonizations
\cite{machartur1967,ricklefs1987}. At a global scale, originations
and extinctions are controlled by mechanisms acting on
macroevolutionary timescales.

From a theoretical viewpoint, the simplest baseline model for population dynamics
is a random walk without drift, according to which the
abundance of a species in a geographic region has the same
probability of increasing or decreasing by one individual at every
time step. According to this scheme, local extinction is
equivalent to a random walker's first passage to zero, and thus
the resulting persistence-time distribution has a power-law decay with
exponent $3/2$~\cite{chandrasekar1943}.

A more realistic
description can be achieved by accounting for basic ecological
processes like birth, death, migration and speciation in neutral
\cite{UNTB,volkov2003,alonso2006,muneepeerakul2008} mean field
schemes, as follows. Consider a community of $N$ individuals
belonging to different species. At every time step a randomly
selected individual dies and space or resources are freed up
for colonization. With probability $\nu$ the site is taken by an
individual of a species not currently present in the system; $\nu$
is equivalent to a per-birth diversification rate and it accounts
for both speciation and immigration from surrounding communities.
With the residual probability $1-\nu$ the died individual is
replaced by one offspring of an individual randomly sampled within
the community~\cite{durett1996,chave2002}. As such the probability
of colonization by a species depends solely on its relative
abundance in the community.  The asymptotic behavior of the resulting persistence-time distribution (i.e. $p_{\tau}(t)$)
exhibits a power-law scaling limited by an exponential cut-off:
\begin{equation}
p_{\tau}(t) \ \propto \ t^{-\alpha}e^{-\nu t},\label{eq1}
\end{equation}
with exponent $\alpha=2$~\cite{pigolotti2005}. In Eq.~\ref{eq1},
time is expressed in generation time units~\cite{UNTB}, i.e. it
has been rescaled such that the birth rate is equal to one.
Notably, in the mean field scheme the probability distribution $p_{\tau}(t)$
depends solely on the diversification rate which accounts for
speciation and migration processes and imposes a characteristic
timescale $1/\nu$ for local extinctions. While per-birth
speciation rates are not expected to vary with the spatial scale
of analysis, per-birth immigration rates are argued to decrease as
the spatial scale increases. In fact, the possible sources of
migration (chiefly dependent on the geometrical properties of the
boundary and the nature of dispersal processes) are argued to
scale sub-linearly with the community size~\cite{chilsom2009},
which in turn is typically linearly proportional to geographic
area~\cite{machartur1967,brown1995}. As continental scales are
approached, migration processes (almost) vanish and the
diversification rate ultimately reflects only the speciation rate.

From an empirical viewpoint, species and genera persistence times deducted from fossil record data have
been suggested to follow either power-law (with non-trivial
exponents in the range $1.5-2$~\cite{sneppen1995,sole1996,newman1999}) or exponential
distributions~\cite{vanvalen1973,sole1996}. It has been argued,
however, that data quality, in particular for species, precludes
a critical assessment~\cite{pigolotti2005}. Also, local analyses of
species persistence over ecological timescales suggest power-law distributions with non trivial exponents
\cite{keitt1998}.

In what follows we shall provide evidence for power-law behavior,
either empirically or from a broad spectrum of theoretical
derivations. Implications on emerging macroecological patterns will be examined,
with special attention to possible biogeographical effects.

\section{Empirical Persistence-Times Distributions}

We empirically characterize species persistence-time distributions by
analyzing two long-term datasets covering very different spatial
scales: i) a 41-year survey of North American breeding birds
\cite{bbs}; and ii) a 38-year inventory of herbaceous plants from
Kansas prairies~\cite{adler2007}.

The North American Breeding Bird Survey consists of a record of
annual abundance of more than $700$ species over the 1966-present
period along more than $5000$ observational routes. The spatial
location of the routes analyzed is shown in Fig. 1. We consider
only routes with a latitude less than $50^\circ$ because  density
of routes with a long surveyed period drastically decreases above
the 50th parallel. Noting that in many regions the survey started
only in $1968$, we discard the first two years of observations in
order to have simultaneous records for all the regions in the
system. The spatial extent of the observational routes allows us
to analyze species persistence times at different spatial scales. We
consider $20$ different scales of analysis with linearly
increasing values of the square root of the sampled area starting
from $A=10000$ km$^2$ to $A=3.8 \cdot 10^6$ km$^2$. We also
analyze the whole system, which corresponds to an area of $A=7.8
\cdot 10^6$ km$^2$. For every scale of analysis $A$ we consider
several overlapping square cells of area $A$ inside the system. A
three-dimensional presence-absence matrix $P$ is thus built. Each
element~$p_{stc}$ of the matrix is equal to $1$ if species~$s$ is
observed during year~$t$ in at least one of the observational
routes comprised in cell~$c$, otherwise $p_{stc} = 0$. For every scale of analysis
we discard the cells that (i) do not have a continuous record for
the whole period (41~years) or (ii) have more than $5\%$ of their
area falling outside the system. For every cell and every species
we measure persistence times from presence-absence time series derived
from the second dimension of matrix~$P$. Persistence time is defined as
the length of a contiguous sequence of ones in the time series.
For every scale of analysis we consider all the measured persistence times
regardless of the species they belong to and the cell where they
are measured. The effect of possible imperfect detection of species \cite{ajara04} on measured persistence times has also been investigated (see Supporting Information (SI)).

The herbaceous plant dataset~\cite{adler2007} comprises a series
of $51$ quadrats of $1$ m$^2$ from mixed Kansas grass prairies
where all individual plants were mapped every year from $1932$ to
$1972$. In order to meet the data quality standard required for
our analysis as discussed above for the breeding bird data, we
discard $10$ quadrats and the first three years of observations.
Due to the limited number of observational plots in the herbaceous
plant dataset we limit our analysis to quadrat spatial scale $A=1$
m$^2$. Analogously to the previous case, we reconstruct the matrix
$P$ from presence-absence data for every species, year and
quadrat.

Note that, when dealing with empirical survey data, the effect of
the finiteness of the observational time window on the measured
species persistence times must be properly taken into account. To this
end, we have developed new tools to extend the inference of
persistence-time distributions for periods longer than the observational
window. In particular,
we analytically derive, given the persistence-time probability density
function, the distribution of two additional variables that can
actually be measured from empirical data: i) the persistence times $\tau'$
of species that emerge and go locally extinct within the observed
time window $\Delta T_w$; and and ii) the variable $\tau''$ that
comprises ~$\tau'$ and all the portions of species
persistence times that are partially seen inside the observational time
window but start or/and end outside (Fig.~\ref{fig2}A and Materials and Methods).
 The finiteness of the time window imposes a
cut-off to $p_{\tau'}(t)$. On the contrary $p_{\tau''}(t)$ has an
atom of probability in $t=\Delta T_w$ corresponding to the
fraction of species that are always present during the
observational time. By matching analytical and observational
distributions for $p_{\tau'}(t)$ and $p_{\tau''}(t)$, it is
possible to infer the persistence-time distribution $p_{\tau}(t)$.
 The scaling exponent and the diversification rate for the
herbaceous plant persistence-time distributions have been determined with
a simultaneous nonlinear fit of observational and analytical
$p_{\tau'}(t)$ and $p_{\tau''}(t)$. Confidence intervals are equal
to the standard error of the fit. For breeding birds, we repeat
the nonlinear fit for different spatial scales of analysis. The
reported scaling exponent and the confidence interval have been
obtained by averaging results across spatial scales.

Remarkably, the persistence times of breeding birds at different spatial
scales of analysis and of herbaceous plants  prove to be best
fitted by a power-law distribution with an exponent $\alpha =1.83
\pm 0.02$ and $\alpha =1.78 \pm 0.08$, respectively
(Fig.~\ref{fig2}B,C).
It is important to note that both scaling coefficients derived empirically are significantly different from the predictions of the
existing baseline models discussed above (the random-walk persistence time yields an exact exponent  $\alpha=3/2$;
the mean-field model yields $\alpha=2$).

\section{Theoretical Persistence-Time Distributions}

In this section we provide a theoretical rationale for the universality of the scaling behavior of persistence-time distributions with respect to the topology of the interactions allowed by the environmental matrix. In particular we
 provide evidence on how non-trivial exponents of the type observed
empirically can be reproduced by simple theoretical models once dispersal
limitation and the actual network of spatial connections are taken
into account. We have implemented the neutral game described above
in regular one-, two- and three-dimensional lattices in which
every site represents an individual~\cite{durett1996,chave2002}.
We have also explored the patterns emerging from the application of the
model to dendritic structures mimicking riverine ecosystems where
dispersal processes and ecological organization are constrained by
the network structure. Indeed, many features of riverine ecosystems have been shown to be affected by the connectivity of river networks~\cite{grant2007,iturbe2009}. In particular, river geometry has been studied in relation to extinction risk~\cite{fagan2002}, migration processes~\cite{campos2006}, persistence of amphibian populations~\cite{grant2010}, macroinvertebrate dispersal~\cite{brown2010} and freshwater fish biodiversity~\cite{muneepeerakul2008,bertuzzo2009}.
For general calculations of the topological structure and metric properties relevant to dendritic ecological corridors, we employ the features of Optimal Channel Networks (OCNs)~\cite{rodriguez1992}. They hold fractal characteristics known to closely conform to the scaling of real networks~\cite{rinaldo1992}. Among the advantages of the use of OCNs, one recalls the possibility to fit one such construct into any assigned domain (e.g. a square, Fig.~\ref{fig3}), thus allowing exactly the same size and number of nodes of a two-dimensional lattice to be endowed with altered directionality of connections.
To account for limited dispersal effects, we allow only the offsprings of the
nearest neighbors of the died individual to possibly colonize the
empty site. In the networked landscape the neighborhood of a site
is defined by the closest upstream and downstream sites. Limited
dispersal promotes the clumping in space of species, which enhances
their coexistence and survival probability
\cite{chave2002,kerr2002}. Indeed we find that in all the
considered landscapes, persistence-time distributions still follow a
power-law behavior characterized by smaller, non-trivial scaling
exponents (namely
$\alpha =1.92$ for the 3D,
$\alpha =1.82$ for the 2D,
$\alpha =1.62$ for the OCN,
$\alpha =1.50$ for the 1D landscape, Fig.~\ref{fig3}) limited by an exponential cut-off.
Remarkably, the exponent obtained  via
simulation in a two-dimensional landscape ($\alpha =1.82 \pm
0.01$) is close to those found in both breeding birds and herbaceous plants datasets ($\alpha =1.83
\pm 0.02$ and $\alpha =1.78 \pm 0.08$, respectively).

We also study how persistence-time distributions deducted from the theoretical
model change with dispersal broader than nearest neighbors (see Fig.~S3). As
expected, as long as the mean dispersal distance remains small
with respect to the system size, the distribution eventually ends up scaling as the one
predicted by the nearest-neighbors dispersal. We also relax the neutral
assumption  by implementing an individual-based
competition/survival tradeoff model~\cite{chave2002}. Specifically,
species with higher mortality rates are assumed to hold less
competitive ability to colonize empty
sites~\cite{brown1995,tilman1994}. It is important to note that the
trade-off model also exhibits power-law persistence-time distribution with exponents indeed close to those shown by the neutral
model (see Fig.~S4). Our theoretical results
are thus robust with respect both to changes in the dispersal range
and to relaxations of the neutrality assumptions. This
confirms our expectation that a power-law distribution for species persistence times is
the result of emergent behaviors independent of fine ecological details, thus supporting the neutral assumption that effective interaction strength among species is weak \cite{volkov2009} and does not significantly constrain the dynamics of ecosystems.
We also note that our results are not seen
as a test for the neutrality hypothesis for breeding
birds or herbaceous plants dynamics, but rather as tools to reveal
emerging universal and macroscopic patterns~\cite{sole2002,pueyo2007}.

\section{Discussion and Conclusions}

In the previous section we have established a hierarchy of scaling exponents ranging from the smallest, proper to one-dimensional (1-D) interactions,  to larger values namely for
directional (network-like), 2-D, and 3-D dispersals.  We thus suggest that the coherence of the empirical scalings would stem from the two-dimensional isotropic nature of the environmental matrix available to the ecological processes relevant to both breeding birds and herbaceous plants.

We also suggest that species persistence-time distribution, owing to its robustness and scale-invariant character, is a synthetic descriptor of ecosystem dynamics and of biodiversity. In fact, other key macroecological patterns are intimately related to the persistence-time distribution. A first clear example is the direct link with ecosystem diversity, as explained below. In our framework species emerge as a
point Poisson process with rate $\lambda=\nu N$ and last for a persistence time $\tau$. The mean number of species $S$ in the system at a
given time is therefore $S=\lambda \langle\tau\rangle$
\cite{rodriguez1987} where $\langle\tau\rangle$ is the mean
persistence time. Therefore, the smaller exponents found, say, for
networked environments with respect to two-dimensional ones, imply
longer mean persistence and, in turn, higher diversity. This echoes
recent results suggesting a higher diversity of freshwater versus
marine ray-finned fishes~\cite{aguiar2009,moyle2003}.

Another evidence of the effective way in which species persistence-time distribution can characterize ecosystem diversity
is the link with the species-area relationship, which
characterizes the increase in the observed number of species with
increasing sample area.
The spatial extent of the breeding bird dataset and the tools
developed for the data analysis allow us to study how the persistence-time
distribution depends on the spatial scale of analysis
(Fig.~\ref{fig4}A). As expected, while the scaling exponent
remains the same, the diversification rate $\nu$ decreases with
the geographic area $A$ and is found to closely follow a scaling
relation of the type $\nu \propto A^{-\beta}$, with $\beta=0.84\pm
0.01$ (Fig.~\ref{fig4}B), for a wide range of areas.
This scaling form of the cut-off timescale $1/\nu$
can be related to the species-area relationship. Assuming that the number of individuals
scales isometrically with the sampled geographic area
\cite{machartur1967,brown1995}, i.e. $N\propto A$, and given that
$\langle\tau\rangle = \int t p_\tau (t) dt \propto \nu^{\alpha-2}$
(see SI) one gets:
\begin{equation}
S \;=\;\lambda\; \langle \tau \rangle \;\propto \; A^{1-\beta(\alpha-1)}\;=\;A^{z}.
\end{equation}
The observational values $\beta=0.84 \pm 0.01$ and $\alpha=1.83
\pm 0.02$ give an exponent $z=0.30 \pm 0.02$ which is close to the
species-area relation measured directly on the data for the same
range of areas ($z=0.31 \pm 0.02$, Fig.~\ref{fig4}C). Conversely,
one could have used the observed species-area exponent to infer
the scaling properties of the diversification rate.

Finally, from a conservation perspective, a meaningful assessment of
species' local extinction rates is deemed valuable. We propose the
distribution of the times to local extinction $\tau_e$
(Fig.~\ref{fig2}A) as a tool to quantify the dynamical evolution of the
species assembly currently observed within a given geographic area.
 Mathematically, $\tau_e$ is defined as the time to
local extinction of a species randomly sampled from the system,
regardless of its current abundance. When Eq.~\ref{eq1} holds for
persistence times, the distribution of the times to local extinction $p_{\tau_e}(t)$ is
shown to scale as $p_{\tau_e}(t) \propto t^{1-\alpha}
e^{-\nu t}$ (see Materials and Methods). Therefore, not only do the developed theoretical and
operational tools allow to infer the scaling behavior of persistence times, but also of the time to local extinction even from
relatively short observational windows. Although these patterns cannot
provide information about the behavior of a specific species or of a
particular patch inside the ecosystem considered (e.g. a biodiversity
hot-spot) they can effectively
describe the overall dynamical evolution of the ecosystem diversity. In particular the scaling behavior allows to extrapolate
species persistence-time distributions for wide geographic areas, which are hard to estimate, from measures of persistence on smaller areas, which are, on the contrary, more practical and feasible.
We thus conclude that the biogeographical characters of species
persistence, stemming from the structure of the spatial interaction networks and from local constraints to species emergence rates, add a new
ingredient to a rich literature bearing major implications for the
inventory of life on Earth.

\section{Materials and Methods}
\subsection{Inference of the Persistence-Times Distribution from a Finite
Observational Period}
The exact derivation of the probability distribution of the
variables $\tau'$ and $\tau''$ (Fig.
\ref{fig2}A) follows.
In this theoretical framework, The probability $\nu dt$ of observing a diversification event in a
time step $dt$ is assumed to be a constant, thus species emergence
in the system due to migration or speciation is seen as a uniform
point Poisson process with rate $\lambda=\nu\,N$ (where $N$ is
total number of individuals in the system and $\lambda$ has the
dimension of the inverse of a generation time). We term $t_0$ the
emergence time of a species in the system, and $T_0$ and
$T_f=T_0+\Delta T_w$ the beginning and the end of the
observational time window, respectively. A species emerged at time
$t_0$ will be continuously present in a geographic region for its
persistence time $\tau$ until its local extinction at time $t_0+\tau$.

We first analyze the distribution of $\tau''$, the most complex
case. The variable  $\tau''$ can be expressed as function of the
random variables $\tau$ and $t_0$, which are probabilistically
characterized. We can distinguish four different cases (Fig.
\ref{fig2}A):
\begin{enumerate}
\item  the species emerges and goes locally extinct within the time window;
\item the species emerges during the observations and it is still present at the end of the time window;
\item the species emerges before the beginning of the observations and goes locally extinct within the time window;
\item the species is always present for all the duration of the observations.
\end{enumerate}
\noindent or, mathematically:

\begin{equation}\label{tau_w}
\tau''=\left\{
\begin{array}{ll}
\tau, & \hbox{if $T_0\le t_0\le T_f$ and $t_0+\tau\le T_f$}\\
T_f-t_0, & \hbox{if $T_0 \le t_0\le T_f$ and $t_0+\tau>T_f$} \\
t_0+\tau-T_0, & \hbox{if $\;\;0<t_0<T_0$ and $T_0 \le t_0+\tau \le T_f$} \\
T_f-T_0, & \hbox{if $\;\;0<t_0<T_0$ and $t_0+\tau>T$}
\end{array}
\right.\nonumber
\end{equation}

We express the probability of observing  $\tau''$ conditional on a
persistence time of duration $\tau$ as:

\begin{align}
&p_{\tau''}(t|\tau) = \nonumber\\
&\frac{1}{\mathcal{N}}\bigg(  \langle\delta(\tau-t)\Theta(t_0-T_0)\Theta(T_f-(t_0+\tau))\Theta(T_f-T_0-\tau)\rangle+  \nonumber \\
& + \langle\delta(T_f-t_0-t)\Theta(t_0-T_0)\Theta(T_f-t_0)\Theta(t_0-(T_f-\tau))\rangle +  \nonumber\\[5pt]
& + \langle\delta(t_0+\tau-T_0-t)\Theta(t_0)\Theta(T_f-t_0-\tau)\Theta(T_0-t_0)\Theta(t_0-T_0+\tau)\rangle + \nonumber \\
&+ \langle\delta(T_f-T_0-t)\Theta(t_0)\Theta(T_0-t_0)\Theta(t_0-T_f+\tau)\Theta(\tau-T_f+T_0)\rangle\bigg),\label{p_tau_w_tau}
\end{align}

\noindent where the operator $\langle\cdot\rangle$ is the ensemble average
with respect to the random variable $t_0$,  $\delta(x)$ and
$\Theta(x)$ are the Dirac delta distribution and the Heaviside
function, respectively.  $\mathcal{N}$ is the normalization constant.

When comparing analytical and observational distributions, we
assume that the system is at stationarity and unaffected
by initial conditions, i.e.  $T_0$ is far from the beginning of
the process. Mathematically this is obtained taking the limit
$T_0,T_f\rightarrow+\infty$ with $T_f-T_0=\Delta T_w$. By solving the ensemble averages and by marginalizing with respect to $\tau$, Eq. \ref{p_tau_w_tau} finally takes the form (see SI for a step by step derivation):

\begin{eqnarray}
p_{\tau''}(t)&=&\frac{1}{\mathcal{N}}\bigg((\Delta T_w-t)p_{\tau}(t)\Theta(\Delta T_w-t)\;+\nonumber\\
 &+& \Theta(\Delta T_w-t)\int_{t>0}^{\infty}p_{\tau}(\tau)d\tau\;+\nonumber \\
&+& \Theta(\Delta T_w-t)\int_{t>0}^{\infty}p_{\tau}(\tau)d\tau\;+\nonumber \\
&+& \delta(t-\Delta T_w)\int_{\Delta T_w}^{\infty}(\tau-\Delta T_w)p_{\tau}(\tau)d\tau \bigg)\; , \label{p_tau_w_final}
\end{eqnarray}

\noindent where $\mathcal{N}$ simplifies to:

\begin{equation}\label{nn1}
\mathcal{N}=\Delta T_{w}+ \langle\tau\rangle-2\,\Delta T_w \,P_{\tau}(\Delta T_w)+2 \bigg(\int_{0}^{\Delta T_{w}}(P_{\tau}(t)-t p_{\tau}(t))dt\bigg) ,\;\nonumber
\end{equation}

\noindent with $P_{\tau}(t)=\int_t^{+\infty}p_{\tau}(\tau)d\tau$ being the
exceedance cumulative distribution of the persistence-time probability
density function.

The variable $\tau'$ comprises only the first of the four cases
listed in Eq.~\ref{tau_w}. Thus the probability
distribution $p_{\tau'}(t)$ follows directly from the first term
of Eq.~\ref{p_tau_w_final}

\begin{equation}\label{p_tau'}
 p_{\tau'}(t)\;=\;\frac{1}{\mathcal{N'}}(\Delta T_w-t)p_{\tau}(t)\Theta(\Delta T_w-t),\nonumber
\end{equation}

\noindent where the normalization constant $\mathcal{N'}$ is equal to

\begin{equation}\label{norm_p_tau'}
 \mathcal{N'}\;=\;\int_{0}^{\Delta T_w}(\Delta T_w-\tau)p_{\tau}(\tau)d\tau,\nonumber
\end{equation}

\noindent which completes the derivation.

\subsection{Distribution of Times to Local Extinction}
\begin{figure}
  \includegraphics[height=39pc]{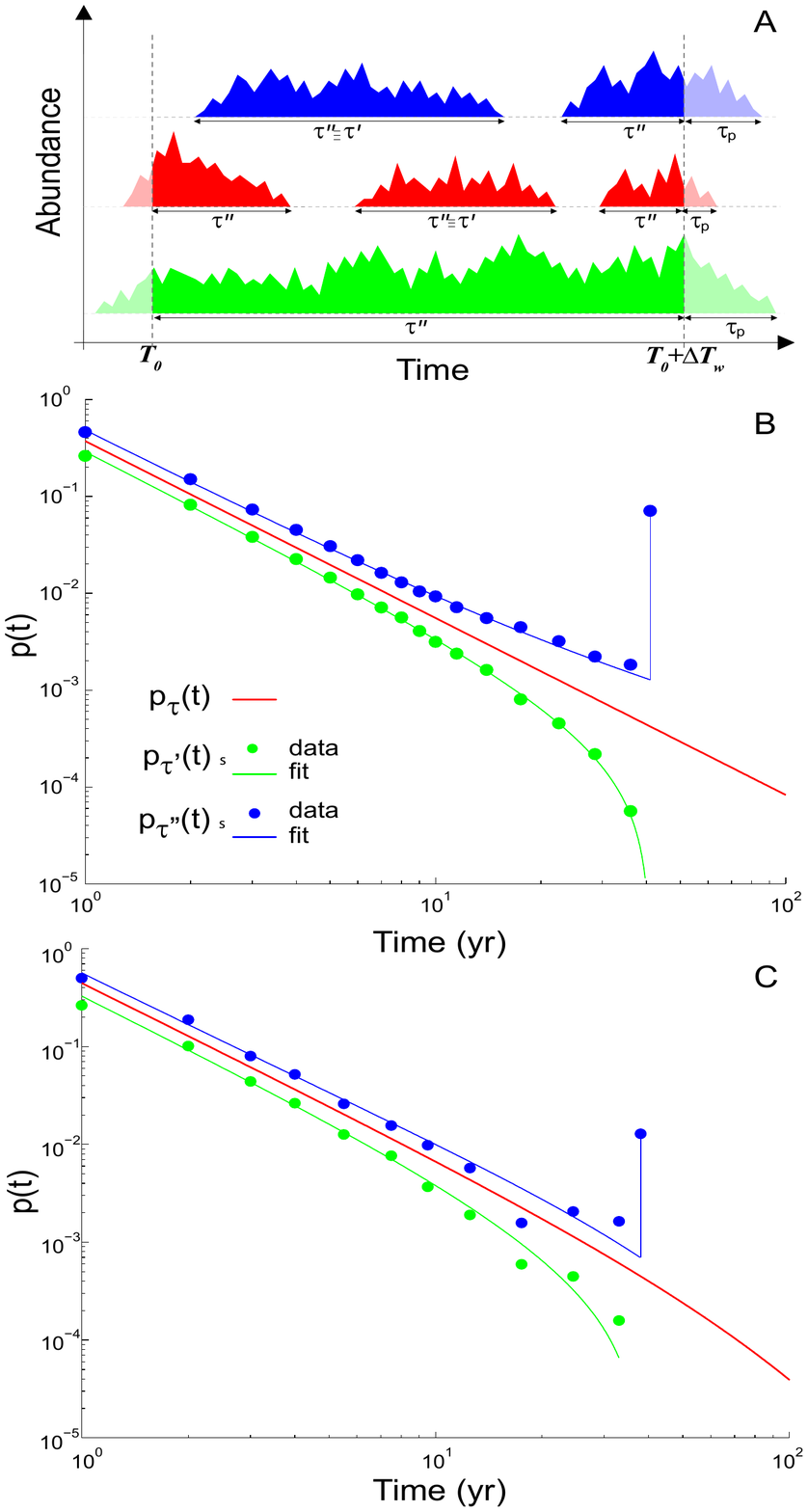}\\
\caption{}\label{fig2}
\end{figure}

\begin{figure}
  \includegraphics[height=29pc]{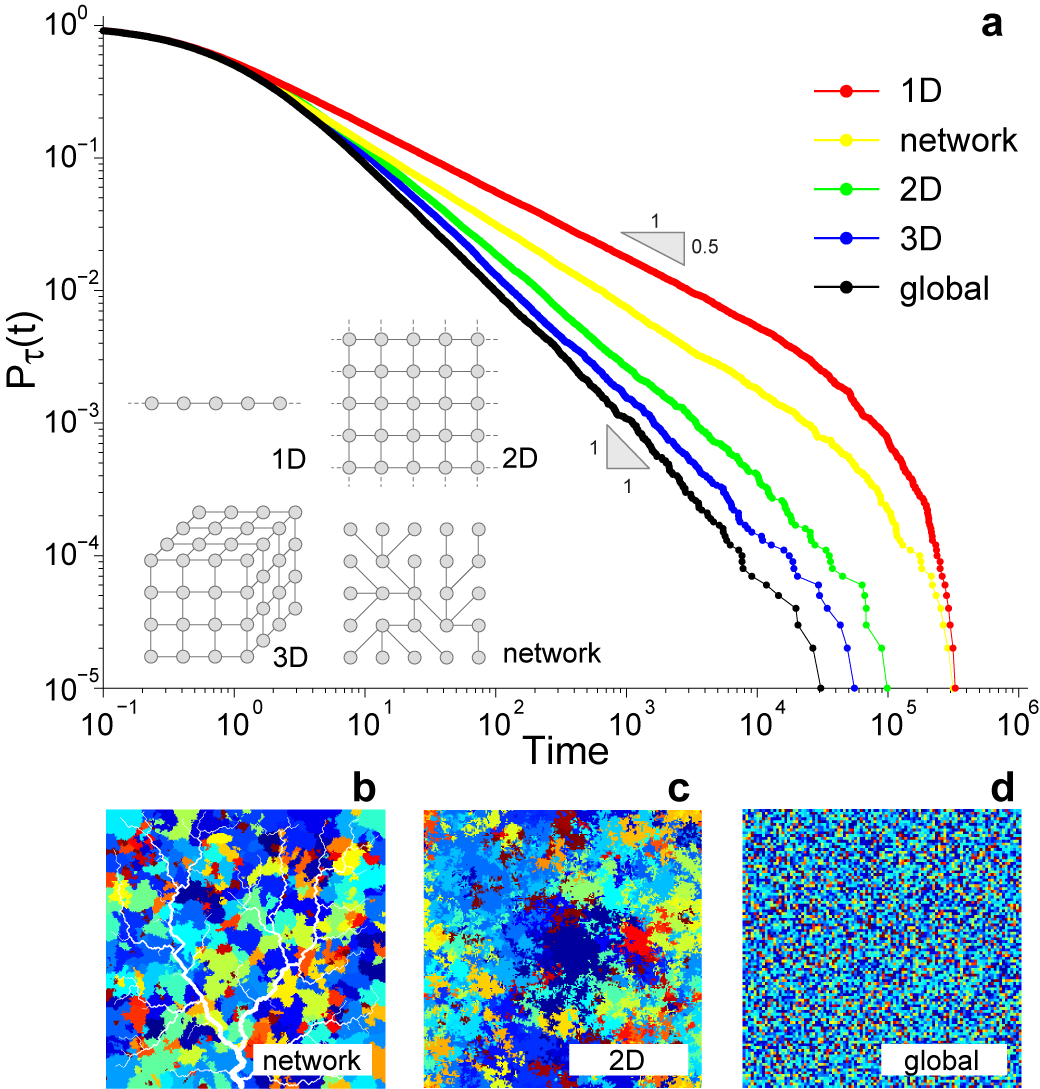}\\
\caption{}\label{fig3}
\end{figure}

\begin{figure}
  \includegraphics[width=29pc]{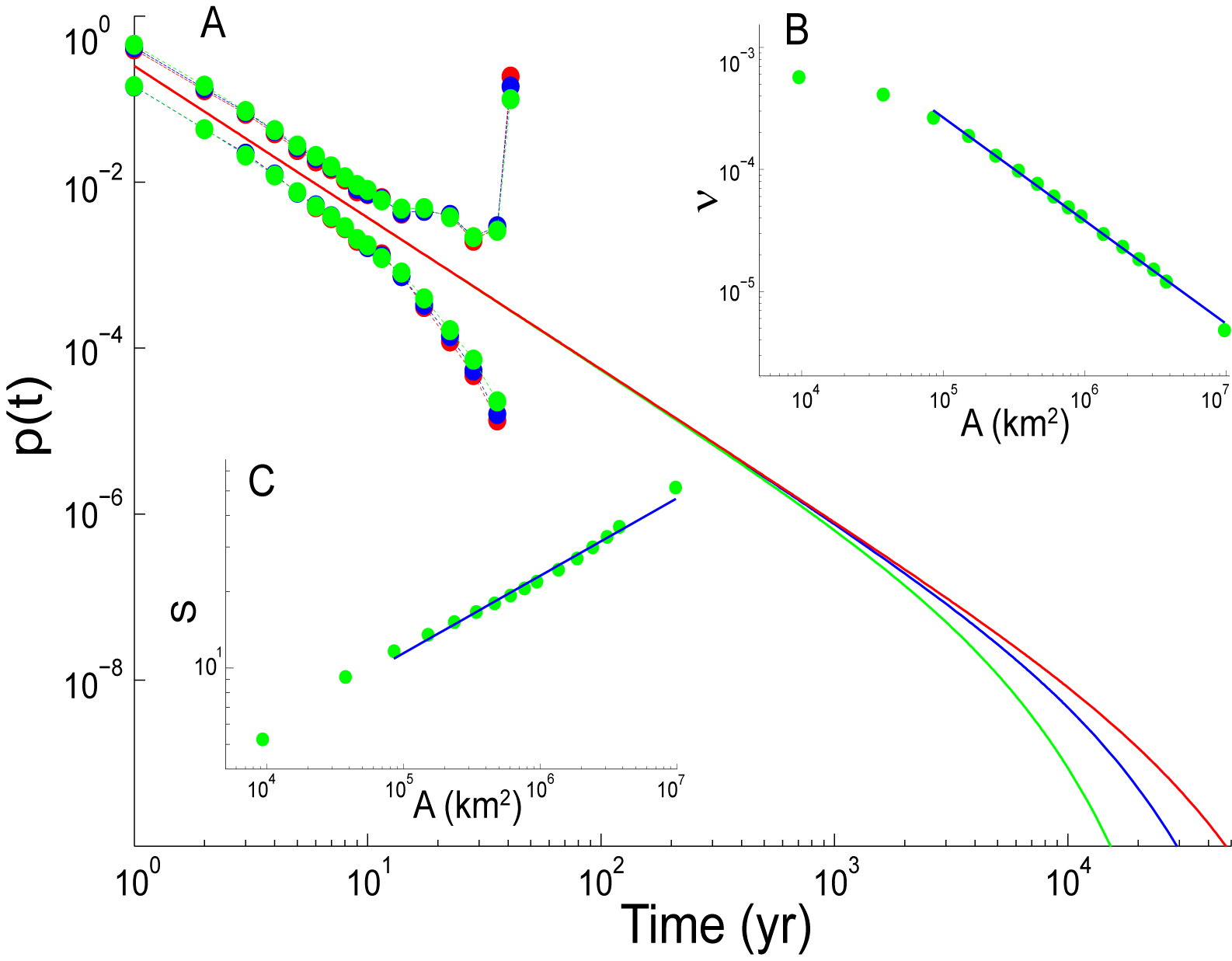}\\
\caption{}
\label{fig4}
\end{figure}

We term $\tau_e$ the time to
local extinction of a species randomly sampled among the observed
assembly at a certain time $T$ (Fig.~\ref{fig2}A).
Analogously to the derivation described above,
we can express $\tau_e$ as:

\begin{equation}\label{tau_s}
    \tau_e=t_0+\tau-T \;\;\;\; \hbox{if $0<t_0<T$ and $t_0+\tau\ge T$.}\nonumber
\end{equation}

We then express the probability distribution of the times to local extinction
conditioned to a persistence time~$\tau$ as:

\begin{equation}\label{p_tau_s_cond}
    p_{\tau_e}(t|\tau)=\frac{1}{\mathcal{C}}\langle \delta(t-(t_0+\tau-T))\Theta(t_0+\tau-T)\Theta(T-t_0)\Theta(t_0)\rangle,\nonumber
\end{equation}

where the constant $\mathcal{C}$ ensures proper
normalization. Solving the ensemble average operators yields

\begin{equation}\label{p_tau_s_cond2}
    p_{\tau_e}(t|\tau)=\frac{1}{\mathcal{C}}\Theta(\tau-T)\Theta(t-\tau+T)\;.\nonumber
\end{equation}

Marginalizing over $\tau$ and considering the system at stationarity ($T\rightarrow+\infty$), we finally obtain

\begin{equation}\label{p_tau_s}
    p_{\tau_e}(t)=\frac{1}{\mathcal{C}}\int_{t}^{\infty}p_{\tau}(\tau)d\tau,
\end{equation}

where $\mathcal{C}$ is simply $\langle \tau \rangle$. Eq.~\ref{p_tau_s} allows to derive the distribution of the times to local extinction given the persistence-time distribution. Particularizing now to the case of persistence-time distributions of the shape $p_{\tau}(t)\propto t^{-\alpha}e^{-\nu t}$, Eq.~\ref{p_tau_s} translates into $p_{\tau_e}(t)\propto t^{1-\alpha}e^{-\nu t}$.\\\\

\emph{acknowledgments}
The authors wish to thank Dr. Stephen Hubbell and two other anonymous referees for their helpful comments and suggestions.
EB, SS, LM and AR gratefully acknowledge the support by the
European Research Council (ERC Advanced Grant RINEC-227612) and by
the Swiss National Science Foundation (project
$200021\_124930/1$). IRI acknowledges the support of the James S.
McDonnell Foundation (grant 220020138). We thank Miguel Munoz for
discussions on the scaling properties  of survival probabilities
and Marino Gatto and Renato Casagrandi for useful comments.

\clearpage
\newpage
\textbf{Figure 1:} Species persistence times. Persistence time $\tau$ within a geographic
region is defined as the time incurred between a species'
emergence and its local extinction. Recurrent colonizations of a
species define different persistence times. The number of
species in the ecosystem as a function of time (gray shaded area) crucially depends on species emergences and persistence times. We analyze two long-term datasets about North America breeding birds~\cite{bbs} and herbaceous plants from Kansas prairies~\cite{adler2007}. The inset shows the observational routes of the Breeding Bird Survey. Aggregating local information comprised in a given geographic area, we reconstruct species presence-absence time-series that allow the estimation of persistence-time distributions.\\

\textbf{Figure 2:} Empirical persistence-time distributions. (a) A schematic representation of the
variables that can be measured from empirical data over a time
window $\Delta T_w$: $\tau'$, persistence times that start and end inside
the observational window, and $\tau''$ , which comprises $\tau'$
and all the portions of persistence times seen inside the time window
that start or/and end outside. Times to local extinction $\tau_e$ are
also presented. (b) Breeding birds and (c), herbaceous plants
probability density function $p(t)$ of $\tau'$ (green), $\tau''$ (blue)
and persistence time $\tau$ (red). Filled circles and solid lines show
observational distributions and fits, respectively. The best
fit is achieved with $p_{\tau}(t)\propto t^{-\alpha}$  with
$\alpha=1.83 \pm 0.02$ and $\alpha=1.78 \pm 0.08$ for breeding
birds and herbaceous plants, respectively. Note that previous estimates~\cite{keitt1998} for (b) are revisited here in the light of the new tools and of a longer dataset. The spatial scale of
analysis is $A= 10,000$ km$^2$ and $\Delta T_w=41$ years for (b) and
A=1 m$^2$ and $\Delta T_w=38$ years for (c). The finiteness of the time
window imposes a cut-off to $p_{\tau'}(t)$ and  an atom of probability in $t=\Delta T_w$ to $p_{\tau''}(t)$, which corresponds to the fraction of species that are always present during the observational time.
$p_{\tau}(t)$ and $p_{\tau'}(t)$ have been shifted in the log-log plot for clarity\\

\textbf{Figure 3:} Persistence-time distributions are dependent on the structure of the spatial interaction networks. (a) Persistence-time
exceedance probabilities $P_\tau(t)$ (probability that species'
persistence times $\tau$ be $\geq t$) for the neutral
individual-based model~\cite{durett1996,chave2002} with nearest-neighbor dispersal
implemented on the different topologies shown in the inset. Note
that in the power-law regime if $p_{\tau}(t)$ scales as
$t^{-\alpha}$, $P_\tau(t) \propto t^{-\alpha+1}$. The scaling
exponent $\alpha$ is equal to $1.50\pm 0.01$ for the one-dimensional lattice (red), $\alpha= 1.62 \pm 0.01$ for the
networked landscape (yellow), $1.82 \pm 0.01$ and $1.92 \pm 0.01$
respectively for the 2D (green) and 3D (blue) lattices. Errors are
estimated through the standard bootstrap method. The persistence-time
distribution for the mean field model (global dispersal) reproduces the
exact value $\alpha=2$ (black curve). For all simulations $\nu=10^{-5}$ and time is expressed in generation time units~\cite{UNTB}. Bottom panels sketch the
color-coded spatial arrangements of species in a networked
landscape (b), in a two-dimensional lattice with nearest neighbor
dispersal (c), and with global dispersal (d).\\

\textbf{Figure 4:} Biogeography of species persistence time. (a) Observational distributions $p_{\tau'}(t)$ and
$p_{\tau''}(t)$ (interpolated solid circles) for the breeding
bird dataset and corresponding fitted persistence-time
distributions $p_{\tau}(t) \propto t^{-\alpha} e^{- \nu t}$ (solid
lines) for different scales of analysis: Area $A=8.5\cdot10^4$
 km$^2$ (green), $A=3.4\cdot10^5$ km$^2$ (blue), $A=9.5\cdot10^5$ km$^2$
(red). $\nu(A)$ provides the cut-off for the distribution,
whose scaling exponent is unaffected by geographic area. Note that
the position of the cut-off of $p_{\tau}(t)$ is inferred
from the estimate of the atom of probability of $p_{\tau''}(t)$ which is more sensitive to the scale of analysis; (b), Scaling of the diversification rate
$\nu$ with the geographic area $\nu \propto A^{-\beta},\;\beta=0.84\pm0.01$;
(c), Empirical species-area relationship (SAR). The
plot shows the mean number of species $S$ found in moving squares of
size $A$. We find $S \propto
A^{z},\;z=0.31\pm0.02$. Slope and confidence interval have been
obtained averaging $41$ SARs, one per year of observation.

\clearpage
\newpage


\clearpage
\newpage
\clearpage
\newpage

\clearpage
\newpage

\clearpage
\newpage

\end{document}